\pdfoutput=1
\documentclass[iop]{emulateapj}
\usepackage{amsmath,amssymb,color,verbatim}
\slugcomment{}
\shorttitle{End-dominated Collapse in Mon~R1 filament}
\shortauthors{N.~K. Bhadari et al.}
\begin{document}
\title {star-forming sites IC 446 and IC 447: an outcome of end-dominated collapse of Monoceros R1 filament}
\author{N.~K. Bhadari\altaffilmark{1,2}, L.~K. Dewangan\altaffilmark{1}, L.~E. Pirogov\altaffilmark{3}, and D.~K. Ojha\altaffilmark{4}}
\email{naval@prl.res.in}
\altaffiltext{1}{Physical Research Laboratory, Navrangpura, Ahmedabad - 380 009, India.}
\altaffiltext{2}{Indian Institute of Technology Gandhinagar Palaj, Gandhinagar - 382355, India.}
\altaffiltext{3}{Institute of Applied Physics of the Russian Academy of Sciences, 46 Ulyanov st., Nizhny Novgorod 603950, Russia.}
\altaffiltext{4}{Department of Astronomy and Astrophysics, Tata Institute of Fundamental Research, Homi Bhabha Road, Mumbai 400 005, India.}

\begin{abstract}
We present an analysis of multi-wavelength observations of Monoceros~R1 (Mon~R1) complex (at d $\sim$760 pc). 
An elongated filament (length $\sim$14~pc, mass $\sim$1465~M$_{\odot}$) is investigated in the complex, which is the most prominent structure 
in the {\it Herschel} column density map. An analysis of the FUGIN $^{12}$CO(1--0) and $^{13}$CO(1--0) line data confirms the existence of the filament 
traced in a velocity range of [$-$5, +1] km s$^{-1}$. The filament is found to host two previously known sites IC~446 and IC~447 at its opposite ends. 
A massive young stellar object (YSO) is embedded in IC~446, while IC~447 contains several massive B-type stars. 
The {\it Herschel} temperature map reveals the extended warm dust emission (at $T_\mathrm{d}\sim$ 15--21 K) toward both the ends of the filament.
The {\it Spitzer} ratio map of 4.5 $\mu$m/3.6 $\mu$m emission suggests the presence of photo-dissociation regions and signature of outflow activity 
toward IC~446 and IC~447. Based on the photometric analysis of point-like sources, clusters of YSOs are traced mainly toward the filament ends. 
The filament is found to be thermally supercritical showing its tendency of fragmentation, which is further confirmed by the detection of a periodic oscillatory pattern (having a period of $\sim$3--4 pc) in the velocity profile of $^{13}$CO. Our outcomes suggest that the fragments distributed toward the filament ends have rapidly collapsed, 
and had formed the known star-forming sites. Overall, the elongated filament in Mon~R1 is a promising sample of the ``end-dominated collapse" scenario, as discussed by Pon et al. (2011, 2012).
\end{abstract}
%
%%%%%%%%%%%%%-----------------------------------------------------------
\section{Introduction}
\label{sec:intro}
After the launch of the {\it Herschel} Space Observatory  \citep{Pilbratt10}, filamentary structures have been commonly found in star-forming regions \citep[e.g.,][]{andre10,andre14}. Filaments having various physical scales, spanning from the scale of the molecular cloud to the scale of protostellar envelope (i.e., $\sim$100--0.1 pc),
have been reported in the literature \citep[e.g.,][]{myers09,tobin10,hacar11,Goodman14,Lu18}, and are often found to nurture star formation activities along their major axis \citep[e.g.,][]{andre10,schneider12,contreras16,dewangan17c}. 
The observational and theoretical studies suggest that filamentary clouds break into fragments of dense clumps and cores, which can further collapse to form young stellar objects (YSOs) as well as massive stars ($>$ 8 M$_{\odot}$) \citep[e.g.,][]{Bally87,bastien83,bastien91,heitsch08,myers09,pon11,pon12,contreras13,clarke15,kainulainen16,dewangan17c,dewangan19}. 
Massive fragments along the filaments have also been observed, where signatures of potential massive protostars are found \citep[e.g.,][and references therein]{Jackson10,Schneider10,Busquet13,Beuther15,dewangan19}. Previous studies suggest that filaments can magnify the infall accretion rates to individual clumps, and thus can form massive stars \citep{Banerjee08,myers09}.
Hence, the role of filaments in star formation has been evident. However, the physical processes that govern the filamentary fragmentation and the formation of massive clumps/cores and clusters of YSOs are still debatable \citep[e.g.,][]{myers09,schneider12,andre14,baug15,contreras16,dewangan15,dewangan16a,dewangan16b,dewangan17a,dewangan17b,dewangan17c,dewangan17d,dewangan17e,dewangan19}. 

Analytical studies of the infinitely long, hydrostatic cylindrical clouds suggest that filaments are gravitationally unstable to axisymmetric perturbations resulting in the fragmentation at equal spacing \citep[e.g.,][]{inutsuka92,inutsuka97,nakamura93}. In the case of long but finite-sized filaments, the local collapse time-scale is observed as a function of relative axial position, and the collapse occurs faster at the ends of the filament compared to its central part \citep[i.e., end-dominated collapse;][]{bastien83,pon11,pon12}. However, observational evidences of numerical studies, in particular ``end-dominated collapse", are a few in the literature \citep[e.g.,][]{Zernickel13,Beuther15,hacar16,kainulainen16,dewangan17c,dewangan19,Yu19}. 
In this context, to observationally assess the existing numerical simulations concerning star-forming filaments, we have selected a nearby star-forming site in Monoceros R1 (hereafter Mon~R1) complex using a multi-scale and multi-wavelength approach.

\begin{figure*}
\epsscale{1}
\plotone{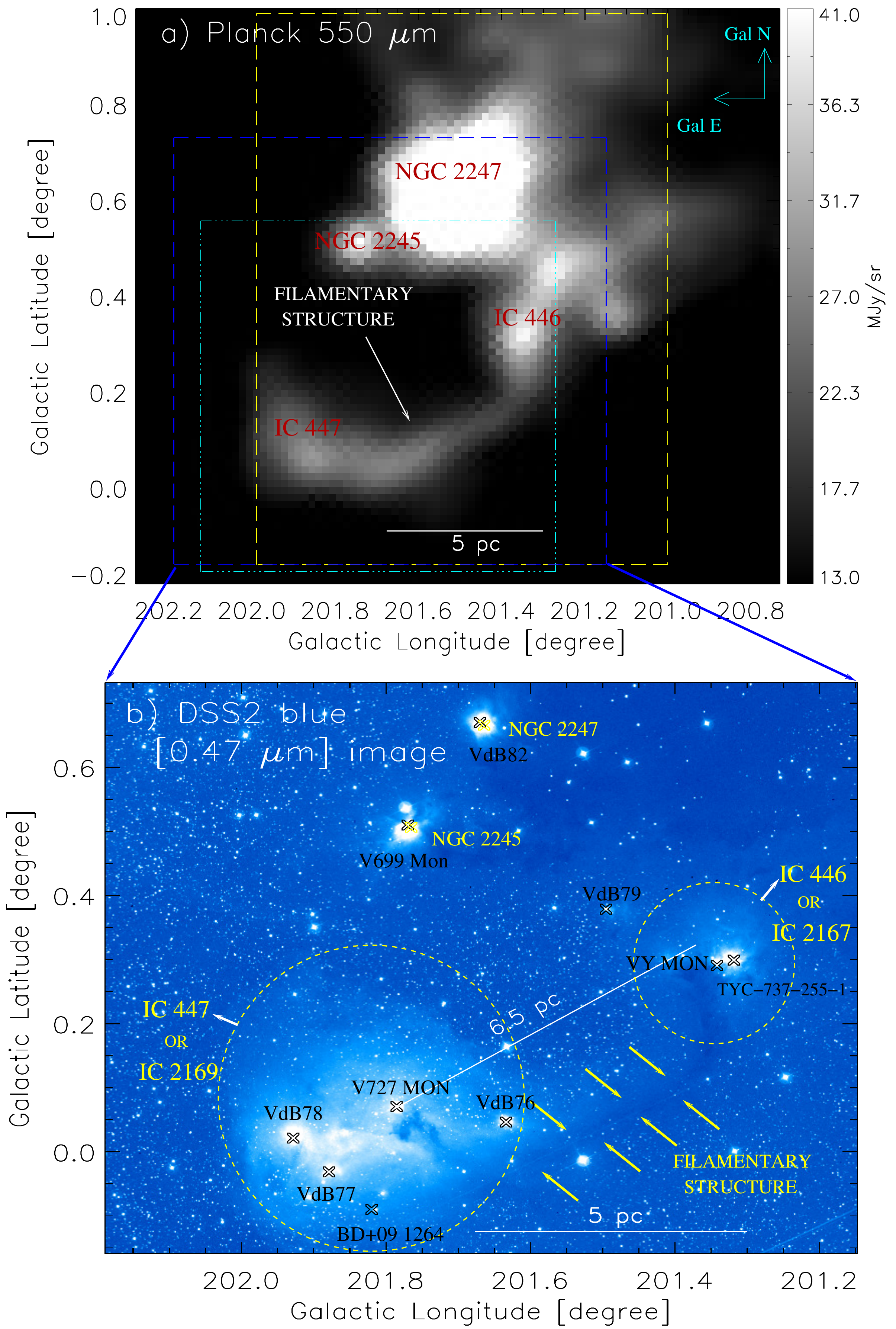}
\caption{The R-association in Mon~R1 complex. a) {\it Planck} continuum image at 550 $\mu$m. A filamentary structure seen in the image is marked by an arrow.
A dashed box (in blue) shows the area presented in Figure~\ref{fg1}b. A dotted-dashed box (in cyan) and a rectangular box (in yellow) represent the areas shown in 
Figures~\ref{fg4} and~\ref{fg2}, respectively. b) A zoomed-in optical image at DSS2 0.47 $\mu$m of Mon~R1 (see a dashed blue box in Figure~\ref{fg1}a). 
The extent of known reflection nebulae IC~446 and IC~447 are shown by dashed circles. 
A dark-filament connecting these nebulae is highlighted by yellow arrows. 
The positions of previously known B-type stars and reflection nebulae (i.e., NGC~2245 and NGC~2247) are 
indicated and labeled by black and yellow color cross symbols, respectively (see Table~\ref{tab3}). 
A scale bar representing 5~pc (at d $\sim$760 pc) is shown in each panel.} 
\label{fg1}
\end{figure*}

Mon R1 is a clustering of reflection nebulae (i.e., R-association), located toward the Galactic plane and allied with the well-studied OB association Mon OB1 \citep{vandenbergh66}. It is situated about 2$\degr$ away from the centre of Mon OB1 \citep[i.e., Cone Nebula NGC 2264; see an inverted C-like structure of Mon R1 association 
(coordinates $\sim$RA: 6$^{\rm h}$ 31$^{\rm m}$ 0$^{\rm s}$, Dec: +10$\degr$ 00$\arcmin$ 00$\arcsec$) in Figure~1 of][]{montillaud19}.
Several distances (e.g., 715 pc \citep{vandenbergh66}, 800 pc \citep{Stahler05}, 900 pc \citep{Oliver96}, 1000 pc \citep{kutner79}) to Mon~R1 are reported in the literature. In this paper, we adopt a heliocentric distance of $\sim$760~pc for Mon~R1 \citep[e.g.,][]{montillaud19}. 
The reflection nebulae NGC~2245, NGC~2247, IC~446, IC~447 and, several early type (B3--B7) stars including VdB 76, VdB 77, VdB 78, are members of the Mon~R1 association \citep{kutner79}. 
Figure~\ref{fg1}a displays the {\it Planck} sub-millimeter image at 550 $\mu$m of R-association, where the locations of NGC~2245, NGC~2247, IC~446, and IC~447 are marked. The direction in the figure is presented in the Galactic coordinates. In Figure~\ref{fg1}b, we show a false-color optical image at Digitized Sky Survey II (hereafter DSS2) 0.47 $\mu$m (size $\sim$1$\degr$.04 $\times$ 0$\degr$.89 or corresponding physical extent $\sim$13.8 pc $\times$ 11.8 pc at d $\sim$760 pc). 
The locations of reflection nebulae along with the previously known bright OB-type stars are also highlighted, and are listed in Table~\ref{tab3}. 
In the optical image, one can clearly see that two sites IC~446 and IC~447 are connected by a curved section of dark nebulosity, which is indicated by arrows in Figure~\ref{fg1}b.
This dark nebulosity appears bright in the {\it Planck} image at 550 $\mu$m, and can be referred to as an elongated filamentary structure. 
In other words, the sites IC~446 (or IC~2167) and IC~447 (or IC~2169) are seen at opposite ends of the filamentary structure. 
A massive YSO, VY Mon is also reported toward the site IC~446 \citep[e.g.,][]{Casey90}.     

Using the molecular CO line observations (resolution $\sim$1$'$.1--2$'$.6), \citet{kutner79} studied the molecular gas toward R-association and Mon OB1. 
The Mon~R1 molecular cloud, traced in a radial velocity (V$_\mathrm{lsr}$) range of [$-$1, 5] km s$^{-1}$, was found to be kinematically distinct from the Mon OB1 molecular complex at [5, 10] km s$^{-1}$. 
A partial ring structure of the Mon~R1 molecular cloud was observed in the molecular map at [$-$1, 5] km s$^{-1}$, which hosts the above mentioned reflection nebulae in its periphery. 
The {\it Planck} continuum image at 550 $\mu$m also shows a semi-ring structure in Figure~\ref{fg1}a. 
However, \citet{kutner79} found at least two velocity components toward Mon~R1 at [$-$1, 1] and [3, 5] km s$^{-1}$ \citep[see Figure~2 in][]{kutner79}. 
The cloud at [3, 5] km s$^{-1}$ was associated with NGC~2245 and NGC~2247, while a semi-circular arc of cloud hosting the sites IC~446 and IC~447 
was reported in a velocity range of [$-$1, 1] km s$^{-1}$ \citep[see Figure~2 in][]{kutner79}. The elongated morphology of the cloud component at [$-$1, 1] km s$^{-1}$ is also observed in other wavelengths; thus we prefer to categorize it as a filament. 
\citet{kutner79} proposed that a previous energetic event, such as expanding H\,{\sc ii} regions, stellar winds, or SN blast waves, was responsible for the existence of the partial ring structure in the Mon~R1 molecular cloud at [$-$1, 5] km s$^{-1}$. They also pointed out that the physical process, which governs the formation of observed OB-stars toward the region, might be different. In this connection, we observationally try to resolve the ambiguity of physical process governing the star formation activity toward the filamentary cloud at [$-$1, 1] km s$^{-1}$.
To understand the physical mechanism(s) of star formation in Mon~R1 cloud, we have examined the distribution of molecular gas, ionized emission, dust (i.e., warm and cold) emission, and YSOs. 

Based on the previous studies, we find that Mon~R1 is a relatively nearby star-forming complex hosting the elongated filamentary structure 
as well as massive B-type stars, making Mon~R1 as an important target site for probing star formation processes. The {\it Herschel} temperature and column density maps (resolution $\sim$12$''$)  are utilized to study the distribution of dust temperature and column density, while high resolution FOREST Unbiased Galactic plane Imaging survey with the Nobeyama 45-m 
telescope \citep[FUGIN;][]{umemoto17} molecular line data (resolution $\sim$20$''$) are employed to examine the gas flow toward the Mon~R1 molecular cloud. 
Furthermore, the observational findings derived in this work have been used to assess the existing theoretical models related to star-forming filaments. 

Following the introduction in this section, we present the adopted data sets in Section~\ref{sec:obser}. The observational outcomes are presented in Section~\ref{sec:data}.
In Section~\ref{sec:disc}, we discuss our observational results against the existing theoretical models. Finally, Section~\ref{sec:conc} gives the main conclusions of this paper.
%%%%%%%%%%%%%%%%%%%%--------------------------------------
%
\begin{table*}
\setlength{\tabcolsep}{0.15in}
\centering 
\caption{List of bright stars known toward Mon~R1 complex (see Figure~\ref{fg1}b). The positions and spectral type of stars are given in the table.} 
\label{tab3}
\begin{tabular}{lcccccccc||cccccccc}
\hline 
\hline
ID                       & Name                            &  {\it l}    &  {\it b}     & Spectral-type & Association       & References       \\   
                         &                                 & (degree)    &  (degree)    &               &                    &                  \\ 
 \hline   
	1    &  TYC 737-255-1	           &    201.32	 &   +00.30	&     B2.5V D  &  IC~446   & --                \\
	2    &   VY Mon &    201.34   &   +00.29     &     A5 Vep   &  IC~446   & \citet{Mora01} \\         
 	3   &	 V727 Mon	           &    201.79	 &   +00.07	&     B8V D    &  IC~447	   &--              \\ 
	4   &	 VdB 76		           &    201.63	 &   +00.05	&     B7IIIp   &  IC~447   & \citet{Racine68}\\
	5   &	 VdB 77		           &    201.88 	 &   $-$00.03 	&     B5III D  &  IC~447   & \citet{Racine68}\\
	6   &	 VdB 78		           &    201.93	 &   +00.02 	&     B3 E     & IC~447	   & \citet{Cannon93}\\
	7   &	 BD+09 1264  	           &    201.82   &   $-$00.09   &     OB E     & IC~447	   & \citet{Nassau65}\\
	8   &   V699 Mon	                   &    201.77	 &   +00.51	&     B7IIne C & NGC~2245	   &\citet{Herbst82}\\
	9    &   VdB 82		           &    201.67	 &   +00.67	&     B6ep D   & NGC~2247	   &\citet{Herbst82}\\
   
\hline          
\end{tabular}
\end{table*}
\section{Data and analysis}
\label{sec:obser}
Figure~\ref{fg1}a displays the target area (central coordinates {\it l} = 201$\degr$.5, {\it b} = 0$\degr$.5) of the present study. 
The selected target area has an angular extent of $\sim$1$\degr$.5 $\times$ 1$\degr$.4, which corresponds to a physical scale of $\sim$20.3 pc $\times$ 18.1 pc (at d $\sim$760 pc). 
Various publicly available multi-wavelength and multi-scale data-sets have been explored in this paper, which are listed in Table~\ref{surveytable}.
In the direction of our selected target area, we retrieved $^{12}$CO(J =1$-$0) and $^{13}$CO(J =1$-$0) line data from the FUGIN survey, which are calibrated in main 
beam temperature \citep[$T_\mathrm{mb}$, see][]{umemoto17}. The typical RMS noise level\footnote[1]{https://nro-fugin.github.io/status/} ($T_\mathrm{mb}$) is $\sim$1.5~K and $\sim$0.7~K for $^{12}$CO and $^{13}$CO lines, respectively \citep{umemoto17}. The FUGIN survey provides the data with a velocity resolution of 1.3 km s$^{-1}$. In order to improve sensitivities, we have smoothened each FUGIN molecular line data cube with a Gaussian function having half-power beamwidth of 3 pixels.

The {\it Herschel} temperature and column density ($N(\mathrm H_2)$) maps (resolution $\sim$12$''$) were obtained from a publicly accessed website\footnote[2]{http://www.astro.cardiff.ac.uk/research/ViaLactea/}, and  were generated for the {\it EU-funded ViaLactea project} \citep{molinari10b}. To produce these maps, the Bayesian algorithm, {\it Point Process Mapping} (PPMAP) method \citep{marsh15,marsh17}, was applied on the {\it Herschel} images at 70, 160, 250, 350, and 500 $\mu$m. 
 \begin{table*}
 % \tiny
\setlength{\tabcolsep}{0.05in}
\centering
\caption{List of multi-wavelength surveys used in this paper.}
\label{surveytable}
\begin{tabular}{lcccr}
\hline 
  Survey  &  band(s)      &  Resolution         &  Reference \\   
\hline
\hline 
NRAO VLA Sky Survey (NVSS) &21 cm 	&$\sim$45$\arcsec$	&\citet{condon98}\\

FUGIN survey   & $^{12}$CO(J = 1--0),$^{13}$CO(J = 1--0) & $\sim$20$\arcsec$, $\sim$21$\arcsec$        &\citet{umemoto17}\\

{\it Planck} Survey	&550 $\mu$m &4$\arcmin$.8 &\citet{planck14}\\
{\it Herschel} Infrared Galactic Plane Survey (Hi-GAL)                              &70, 160, 250, 350, 500 $\mu$m                     & 5$\arcsec$.8, 12$\arcsec$, 18$\arcsec$, 25$\arcsec$, 37$\arcsec$    &\citet{molinari10a}\\
Wide Field Infrared Survey Explorer (WISE) & 22 $\mu$m                   & $\sim$12$''$           &\citet{wright10}\\ 

Warm-{\it Spitzer} GLIMPSE360 Survey      &3.6, 4.5 $\mu$m                   & $\sim$2$''$           &\citet{benjamin03}\\
UKIRT near-infrared Galactic Plane Survey (GPS)                                                 &1.25--2.2 $\mu$m                   &$\sim$0$''$.8           &\citet{lawrence07}\\ 
Two Micron All Sky Survey (2MASS)                                                 &1.25--2.2 $\mu$m                  & $\sim$2$''$.5          &\citet{skrutskie06}\\
Digitized Sky Survey II (DSS2) &0.47 $\mu$m	& $\sim$2$''$ &\citet{mclean00}\\
\hline          
\end{tabular}
\end{table*}
\section{Results}
\label{sec:data}
\subsection{{\it Herschel} column density and temperature maps}
\label{sec:tmcd}
\begin{figure*}
\epsscale{1}
\plotone{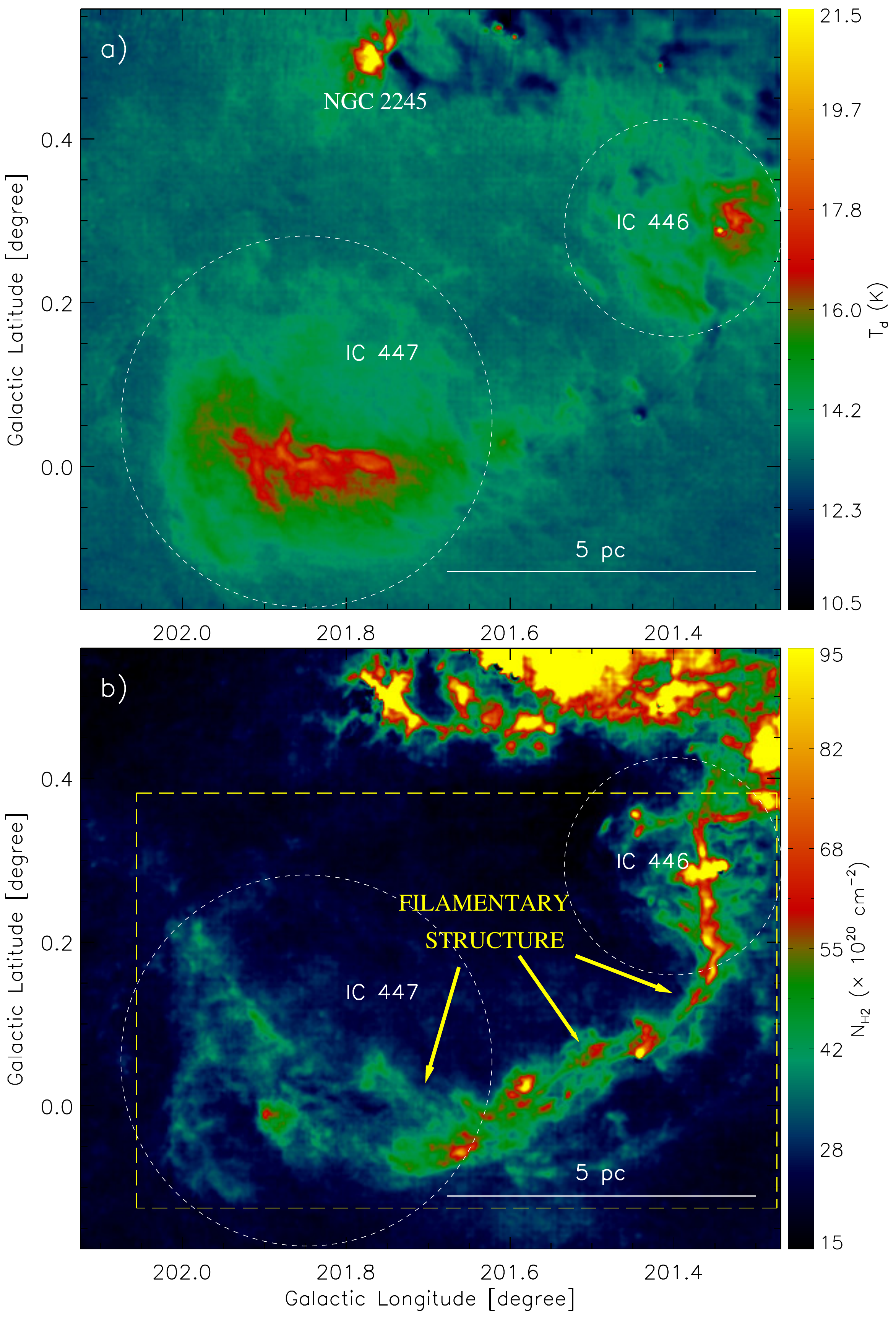}
\caption{a) {\it Herschel} temperature map of the region containing the filament (see a dotted-dashed cyan box in Figure~\ref{fg1}a). 
b) {\it Herschel} column density map. The identification of {\it Herschel} dust clumps is carried out in the area shown by a dashed yellow box (see Figure~\ref{fg5}).
In each panel, broken circles indicate the extent of IC~446 and IC~447. A scale bar corresponding to 5~pc is shown in both panels.} 
\label{fg4}
\end{figure*}

\begin{figure*}
\epsscale{0.8}
\plotone{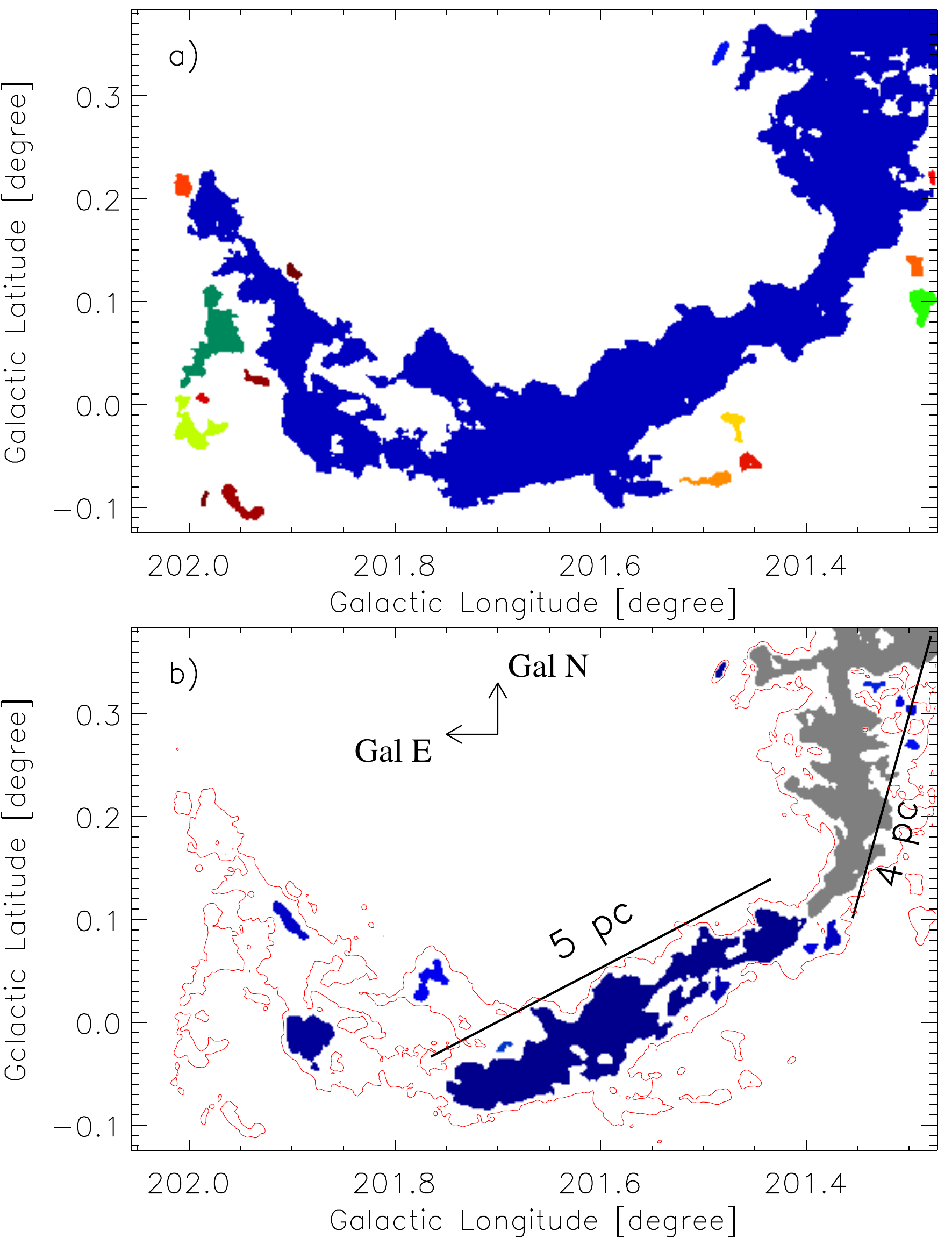}
\caption{a) An elongated filament is depicted in the column density map with the $N(\mathrm H_2)$ contour level of 3.2 $\times$ 10$^{21}$ cm$^{-2}$. b) Two sub-filaments (central in blue, and western in gray) are identified in the direction of the elongated filament using the $N(\mathrm H_2)$ contour level of 4.1 $\times$ 10$^{21}$ cm$^{-2}$. 
The projected length of each sub-filament is indicated by a scale bar. The $N(\mathrm H_2)$ contour (in red) is also overplotted to show the elongated filament as shown in Figure~\ref{fg5}a.}
\label{fg5}
\end{figure*}

Figures~\ref{fg1}a and~\ref{fg1}b have enabled us to infer the existence of the elongated filamentary structure. In order to further examine the filament, in Figures~\ref{fg4}a and~\ref{fg4}b, we display the {\it Herschel} temperature and column density ($N(\mathrm H_2)$) maps of the target region (see a dashed cyan box in Figure~\ref{fg1}a), respectively. 
The {\it Herschel} temperature map shows the extended warm dust emission (at T$_\mathrm{d}$ $\sim$15--21~K) toward the sites IC~446, IC~447, and NGC~2245. 
The elongated filament is revealed as the most prominent feature in the {\it Herschel} column density map and is traced using the $N(\mathrm H_2)$ contour level of 3.2 $\times$ 10$^{21}$ cm$^{-2}$. It is not a straight filament, but shows a curved geometry. The column density map also exhibits the presence of several high column density (i.e., 6--9 $\times$ 10$^{21}$ cm$^{-2}$) regions toward the filament. 
The boundary of the elongated filament (length $\sim$14~pc) is shown in Figure~\ref{fg5}a and is identified using the IDL {\it clumpfind} algorithm \citep{williams94}. 
The {\it clumpfind} algorithm decomposes 2- and 3-dimensional data into disjoint clumps of emission. The algorithm requires contouring of data with a multiple of RMS noise and then finds positions of peak emission which correspond to the clumps, and further follows them down to the lower intensities \citep{williams94}. In other words, one needs the lowest contour level, below which data is rejected, and the interval between contours as input parameters for the {\it clumpfind} algorithm.

Considering the curved morphology of the filament, it can be divided into three parts, which are referred to as eastern, central, and western parts. 
The eastern one, toward the site IC~447, is more diffuse with a single column density peak, while 
the central part (length $\sim$5 pc) and the western one toward the site IC~446 (length $\sim$2--4 pc) are more dense and have several column density peaks. 
In the direction of the elongated filament, Figure~\ref{fg5}b displays the boundaries of two sub-filaments identified using the IDL {\it clumpfind} algorithm. 
Here, the $N(\mathrm H_2)$ contour level of 4.1$\times$ 10$^{21}$ cm$^{-2}$ was used as an input for the {\it clumpfind} algorithm. These sub-filaments are seen toward the central and western parts of the elongated filament as highlighted in Figure~\ref{fg5}a. 
These sub-filaments appear linear, but are not colinear to each other.

To calculate the total mass of the elongated filament (length $\sim$14~pc) and sub-filaments, we use the following relation as given in \citet{dewangan17c}
\begin{equation}
	M_{\rm clump}= \mu_{\rm H_{2}}~m_{\rm H}~ a_{\rm pixel}~\Sigma N(\mathrm H_2) 
\label{eq1}
\end{equation}
where, $\mu_{\rm H_{2}}$ is the mean molecular mass (assumed to be 2.8), $a_{\rm pixel}$ is the area subtended by one pixel, $m_{\rm H}$ is the mass of hydrogen atom, and $\Sigma N(\mathrm H_2)$) is the integrated column density over the area of the clump. We have computed the total mass of the elongated filament (M$_{filament}$) to be $\sim$1465 M$_{\odot}$. 
The total mass of the central and western parts of the filament is determined to be $\sim$319 and $\sim$365 M$_{\odot}$, respectively. 
The importance of these results is discussed in Section~\ref{sec:age}.

In general, the observed line mass of a filament can be calculated as the ratio of M$_{filament}$ to its length. 
In this work, adopting the values of M$_{filament}$ $\sim$1465 M$_{\odot}$ and length $\sim$14 pc, the observed line mass of the filament is determined 
to be M$_\mathrm{l,obs}$ $\sim$105 M$_{\odot}$ pc$^{-1}$. Length of the filament is measured along its major axis with the use of DS9 
software\footnote[3]{http://hea-www.harvard.edu/RD/ds9/}. The line mass of the filament depends upon the ``cos {\it i}" factor \citep[e.g.,][]{kainulainen16}, where ``{\it i}" is the angle between the filament's major axis and the sky plane. Since we do not have the information of ``{\it i}", we assume that the filament lies to the same plane as that of the sky plane. 
Hence, the value of the inclination angle can be taken as zero (i.e., {\it i} = 0), yielding the value of cos {\it i} = 1. Therefore, the observed line mass per unit length can be assumed as an upper limit \citep[e.g.,][]{kainulainen16}.
Additionally, the total mass of the filament (M$_{filament}$ $\sim$1465 M$_{\odot}$) can be underestimated because its small portion (FoV $\sim$8$'$--14$'$) is unavailable in the {\it Herschel} column density map at higher latitude (for comparison, see Figures~\ref{fg5}a and \ref{fg2}c). It is clear from Eq.~\ref{eq1}, the uncertainty in the mass of filament varies linearly with 
the $N(\mathrm H_2)$ uncertainties. 
\subsection{Infrared and Radio view of R-association}
\label{sec:irview}
\begin{figure*}
\epsscale{1}
\plotone{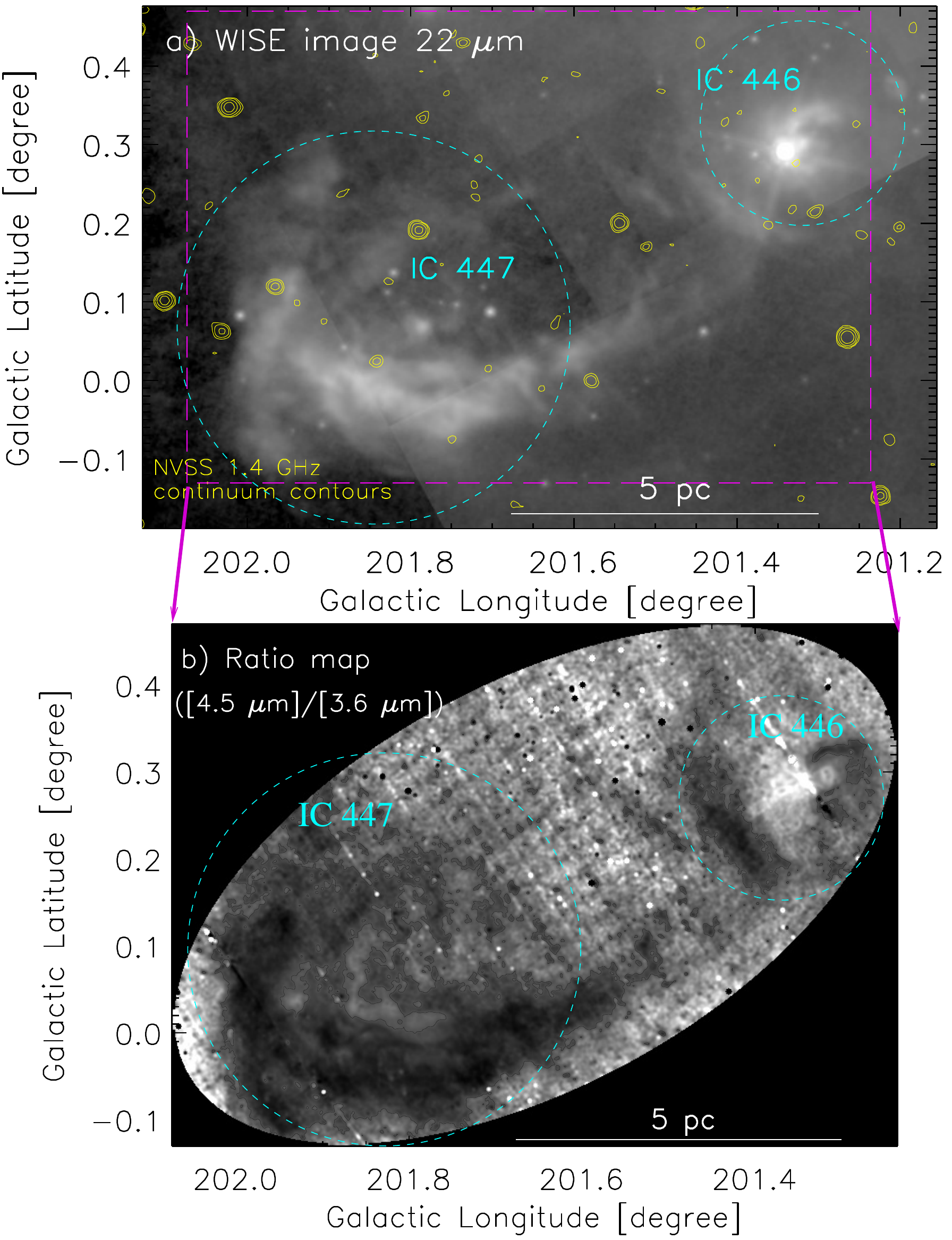}
\caption{a) Overlay of the NVSS radio 1.4 GHz continuum emission contours on the MIR image at {\it WISE} 22 $\mu$m toward the filament. 
The NVSS contours are shown with the levels (3, 5, 10, 20) $\times$ 1$\sigma$, where 1$\sigma$ = 0.45 mJy beam$^{-1}$. A dashed box (in magenta) encompasses the area shown in 
Figure~\ref{fg8}b. b) {\it Spitzer} ratio map of 4.5 $\mu$m/3.6 $\mu$m emission. Broken circles show the extent of IC~446 and IC~447 in each panel. 
The ratio map is smoothened using a Gaussian function with radius of three pixels. A scale bar corresponding to 5 pc is displayed in both panels.} 
\label{fg8}
\end{figure*}

Figure~\ref{fg8}a presents the mid-infrared (MIR) image at WISE 22 $\mu$m of the R-association, which may trace the warm dust emission in our selected target area. 
The WISE image is overlaid with the NVSS 1.4 GHz continuum emission contours \citep[beam size $\sim$45$\arcsec$; 1$\sigma$ $\sim$0.45 mJy beam$^{-1}$;][]{condon98}. 
No extended H{\sc ii} region is found in our selected target area. However, several massive stars have been reported toward the sites IC~447 and IC~446 (see Figure~\ref{fg1}b and also Table~\ref{tab3}), where extended warm dust emission is observed.

We have also carefully examined the {\it Spitzer} 3.6 and 4.5 $\mu$m continuum images to explore the physical environment in Mon~R1. 
In Figure~\ref{fg8}b, we display the {\it Spitzer} ratio map of 4.5 $\mu$m/3.6 $\mu$m emission, which shows bright and dark regions depicting the dominance of 4.5 $\mu$m flux over the 3.6 $\mu$m flux and vice versa. 
It is known that these two {\it Spitzer} images have the same point spread function (PSF). Hence, one can produce the {\it Spitzer} ratio map of 4.5 $\mu$m/3.6 $\mu$m emission, 
which allows us to remove point-like sources as well as continuum emission \citep[e.g.,][]{dewangan17a}. 
The {\it Spitzer} band at 3.6 $\mu$m contains polycyclic aromatic hydrocarbon (PAH) emission at 3.3 $\mu$m, while the molecular hydrogen line ($\nu$ = 0--0 S(9); 4.693 $\mu$m) and 
the Br-$\alpha$ emission (at 4.05 $\mu$m) are covered in the {\it Spitzer} band at 4.5 $\mu$m.
In the ratio map, the bright emission is found toward IC~446, where no NVSS continuum emission (or ionized emission) is detected. 
Hence, the area with bright emission toward IC 446 may be associated with the outflow activity (see also Section~\ref{sec:yso}). 
Furthermore, in the ratio map, the dark or grey regions toward IC~447 display the dominance of 3.3 $\mu$m PAH feature, indicating the presence of photodissociation 
regions \citep[PDRs; see][]{tielens08}. It also favours the impact of massive stars toward IC~447.
\subsection{Kinematics from spectral line data}
\label{sec:moleculargas}
In this section, we present the results derived using the analysis of the FUGIN $^{12}$CO(J =1$-$0) and $^{13}$CO(J =1$-$0) line data.
\subsubsection{Integrated molecular maps}
\label{sec:int1}
\begin{figure*}
\epsscale{1}
\plotone{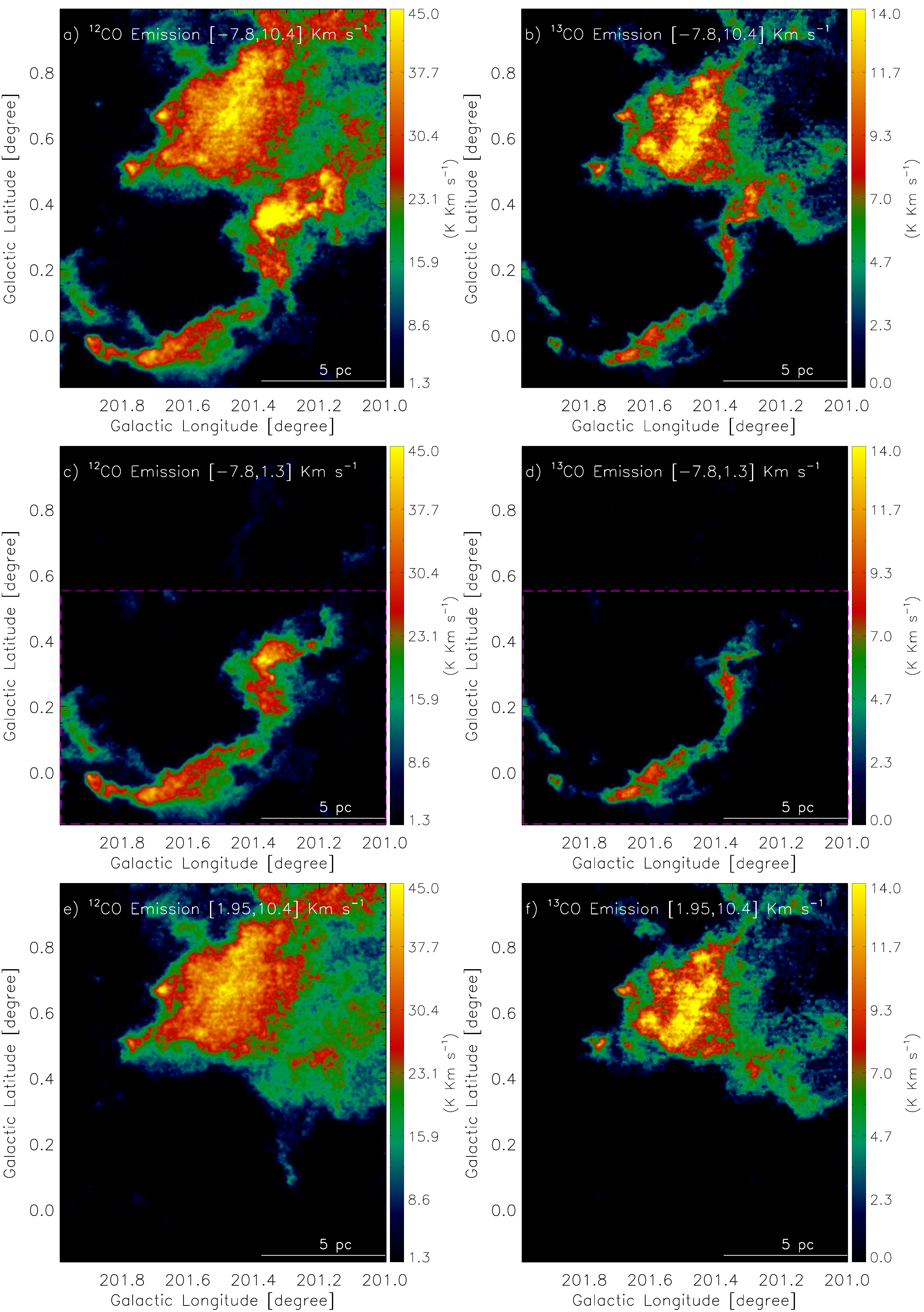}
\caption{Left column: ``a)", ``c)", ``e)" FUGIN $^{12}$CO(1--0) map of intensity (moment-0) in the direction of Mon~R1. 
Right column: ``b)", ``d)", ``f)" FUGIN $^{13}$CO(1--0) map of intensity (moment-0) in the direction of Mon~R1. 
In panels ``a" and ``b", the molecular emission is integrated over a velocity range of [$-$7.8, 10.4] km s$^{-1}$.
In panels ``c" and ``d", the molecular emission is integrated over a velocity range of [$-$7.8, 1.3] km s$^{-1}$.
In panels ``e" and ``f", the molecular emission is integrated over a velocity range of [1.95, 10.4] km s$^{-1}$.
A scale bar corresponding to 5 pc is shown in each panel. In panels ``c" and ``d", the molecular emission is observed mainly toward the filament, which is highlighted by a broken box (in magenta).} 
\label{fg2}
\end{figure*}

In this section, using the FUGIN $^{12}$CO(J = 1--0) and $^{13}$CO(J= 1--0) line data, we present kinematic properties of the molecular cloud(s) associated with our selected target area (see a broken yellow box in Figure~\ref{fg1}a). Figures~\ref{fg2}a--\ref{fg2}f display the CO maps of intensity (moment-0) in the direction of Mon~R1.
In Figures~\ref{fg2}a and~\ref{fg2}b, the $^{12}$CO and $^{13}$CO emissions are integrated over a velocity range of [$-$7.8, 10.4] km s$^{-1}$, respectively. 
The molecular maps at [$-$7.8, 10.4] km s$^{-1}$ show the gas toward NGC~2245, NGC~2247, and the filament containing the sites IC~446 and IC~447. 
Figures~\ref{fg2}c and~\ref{fg2}d present the $^{12}$CO and $^{13}$CO maps integrated over a velocity range of [$-$7.8, 1.3] km s$^{-1}$, respectively. 
In these maps, the molecular emission is observed mainly toward the filament containing the sites IC~446 and IC~447.  
In Figures~\ref{fg2}e and~\ref{fg2}f, the $^{12}$CO and $^{13}$CO emissions are integrated over a velocity range of [1.95, 10.4] km s$^{-1}$, respectively. 
The molecular maps at [1.95, 10.4] km s$^{-1}$ display the molecular emission toward the sites NGC~2245 and NGC~2247. 
As mentioned in the Introduction section, the previous study of the Mon~R1 clouds \citep{kutner79} suggested the presence of two distinct clouds in the velocity range of [$-$1, 1] 
and [3, 5] km s$^{-1}$. However, our analysis is benefited with the availability of high resolution molecular line data compared to the previous study \citep[e.g.,][]{kutner79}. 
In this paper, the analysis of the FUGIN line data also favours the existence of two distinct molecular clouds in the direction of Mon~R1 as discussed by \citet{kutner79}.

\begin{figure*}
\epsscale{1}
\plotone{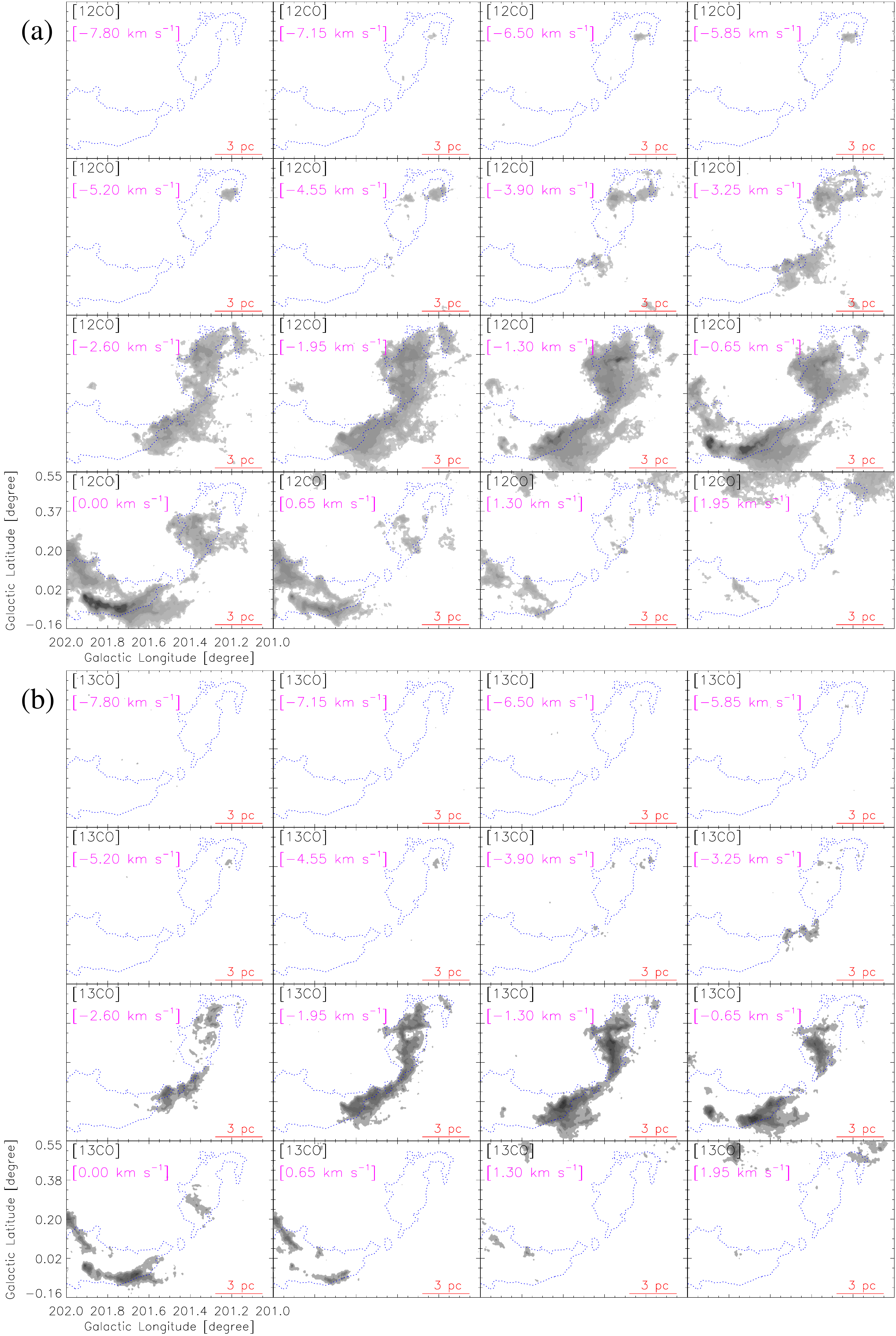}
\caption{a) Velocity channel maps of $^{12}$CO(1--0) toward the region marked by a dashed magenta box in Figures~\ref{fg2}c and~\ref{fg2}d. 
The contour levels are 1.5, 3, 5, 7, 10, 13, 15, 16, 20, and 23 K km s$^{-1}$. 
b) Velocity channel maps of $^{13}$CO(1--0). The contour values are 0.1, 0.5, 1, 1.5, 2, 3, 4, 5, 6, and 7 K km s$^{-1}$. 
The {\it Herschel} 160 $\mu$m continuum emission contour (in blue with a level of 0.02 Jy pixel$^{-1}$ is overplotted in each panel. 
In each panel, the velocity information (in km s$^{-1}$) and a scale bar representing 3 pc are shown.}
\label{fg6}
\end{figure*}

To study the gas flow in the filament, we have selected an area (see a broken box in Figure~\ref{fg2}; extension $\sim$0$\degr$.99 $\times$ 0$\degr$.71 or $\sim$13.14 pc 
$\times$ 9.42 pc at d $\sim$760 pc) toward the target region.
In Figures~\ref{fg6}a and~\ref{fg6}b, the channel maps of $^{12}$CO and $^{13}$CO are presented in a velocity range of [$-$7.8, 1.95] km s$^{-1}$, respectively.
To examine the gas flow in the direction of the filament, we have overlaid the {\it Herschel} 160 $\mu$m continuum emission contour to highlight the extent of the filament. 
The molecular emission is prominently seen in the velocity range of [$-$2.6, $-$0.65] km s$^{-1}$, allowing to trace the boundary of the filament. 
However, in the channel maps of $^{12}$CO and $^{13}$CO, the gas flow toward the filament is observed in a velocity range of [$-$5, +1] km s$^{-1}$.
\subsubsection{Position Velocity ({\it p--v}) diagrams}
\label{sec:pv}
\begin{figure*}
\epsscale{1}
\plotone{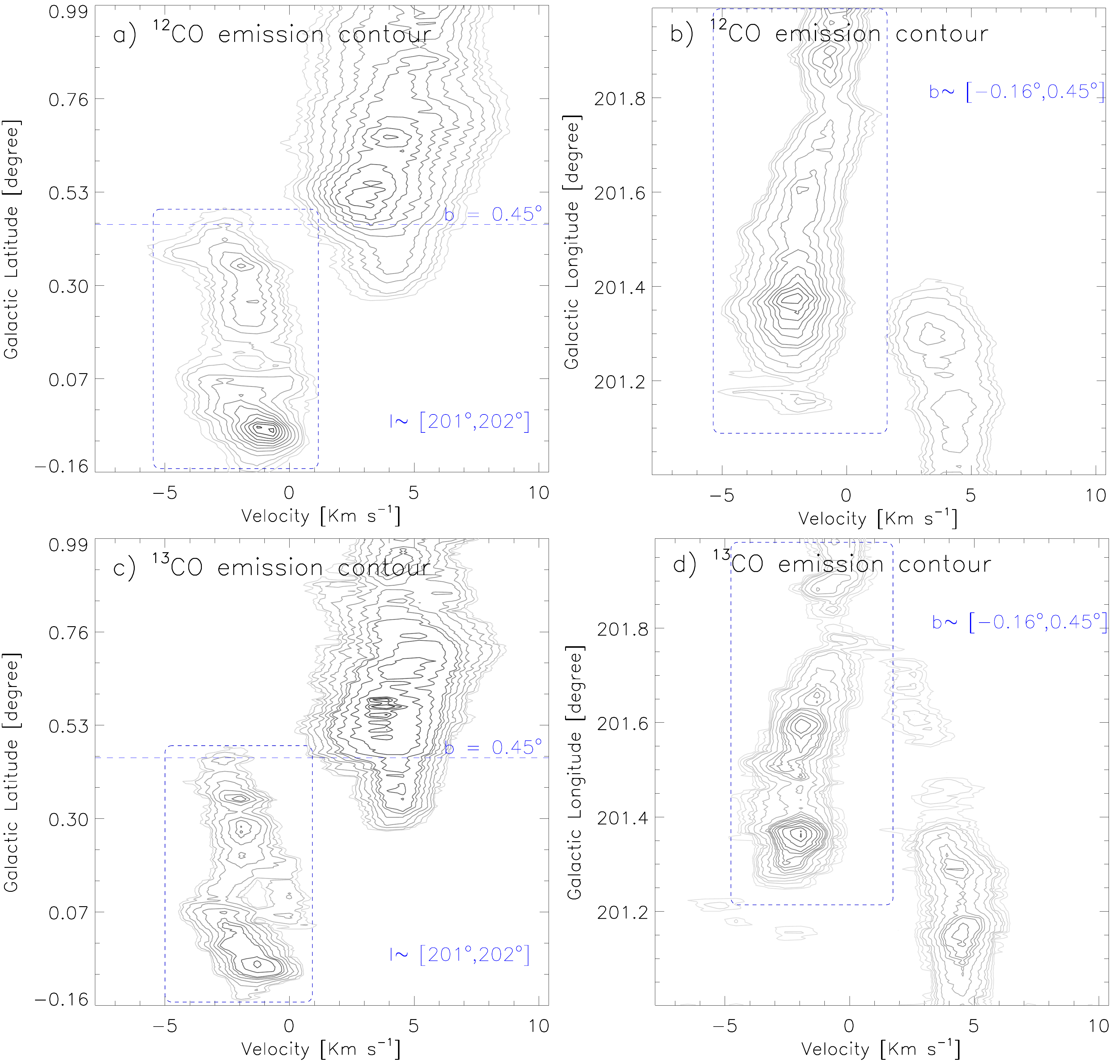}
\caption{a) Latitude-velocity diagram of $^{12}$CO. b) Longitude-velocity diagram of $^{12}$CO.
c) Latitude-velocity diagram of $^{13}$CO. d) Longitude-velocity diagram of $^{12}$CO. 
In the latitude-velocity diagrams (see panels ``a" and ``c"), the molecular emission is integrated over the longitude range 
from 201$\degr$ to 202$\degr$ (see a broken box in Figures~\ref{fg2}c and~\ref{fg2}d). 
In the longitude-velocity diagrams (see panels ``b" and ``d"), the molecular emission is integrated over the latitude range from $-$0$\degr$.16 to 0$\degr$.45. 
In each panel, a dashed rectangular box (in blue) depicts the CO emission toward the filament (see text for more details).} 
\label{fg3}
\end{figure*}

The presence of two clouds is further examined by the analysis of the position-velocity ({\it p--v}) diagrams of molecular gas. 
Figures~\ref{fg3}a and~\ref{fg3}c present the latitude-velocity ({\it b}--{\it v}) diagrams of $^{12}$CO and $^{13}$CO toward the field of view as shown in Figure~\ref{fg2}, respectively. 
In order to extract the {\it b}--{\it v} diagrams, the molecular emission is integrated over the longitude range 
from 201$\degr$ to 202$\degr$.  
In Figures~\ref{fg3}b and~\ref{fg3}d, we show the longitude-velocity ({\it l}--{\it v}) diagrams of $^{12}$CO and $^{13}$CO, respectively.
To obtain the {\it l}--{\it v} diagrams, the molecular emission is integrated over the latitude range from $-$0$\degr$.16 to 0$\degr$.45.

All the position-velocity diagrams show two velocity components toward Mon~R1, which do not seem to be connected in velocity space. 
Hence, we find that the molecular gas toward the filament is traced mainly in a velocity range of [$-$5, +1] km s$^{-1}$ (see a dashed rectangular box in Figure~\ref{fg3}).
These exercises also allow us to infer the exact boundary of the filamentary cloud, suggesting the existence of a single and isolated filament in Mon~R1 (see also Figure~\ref{fg6}).

\subsubsection{Velocity field of the gas}
\label{sec:velstfn}
\begin{figure*}
\epsscale{0.47}
\plotone{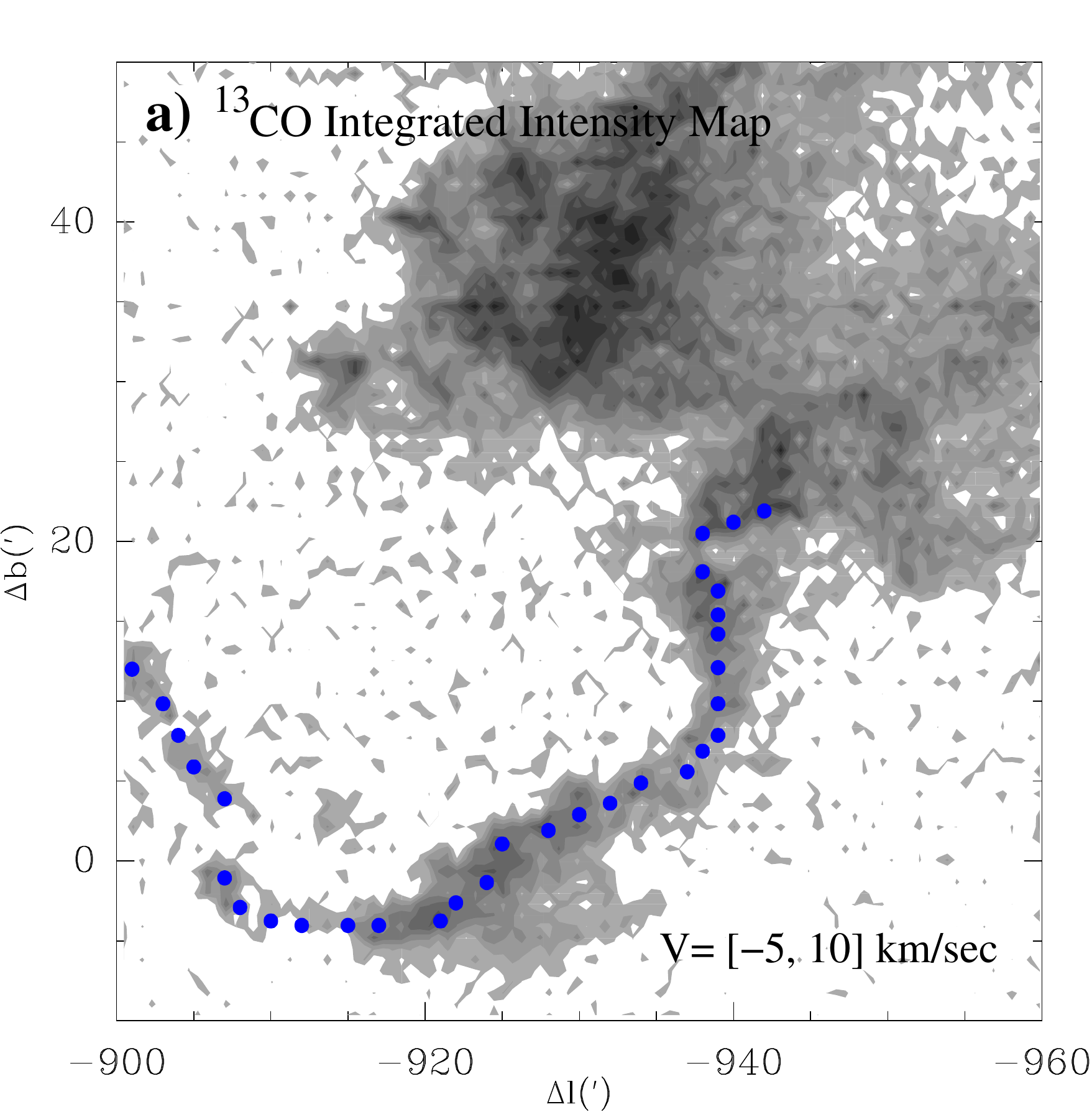}
\epsscale{0.68}
\plotone{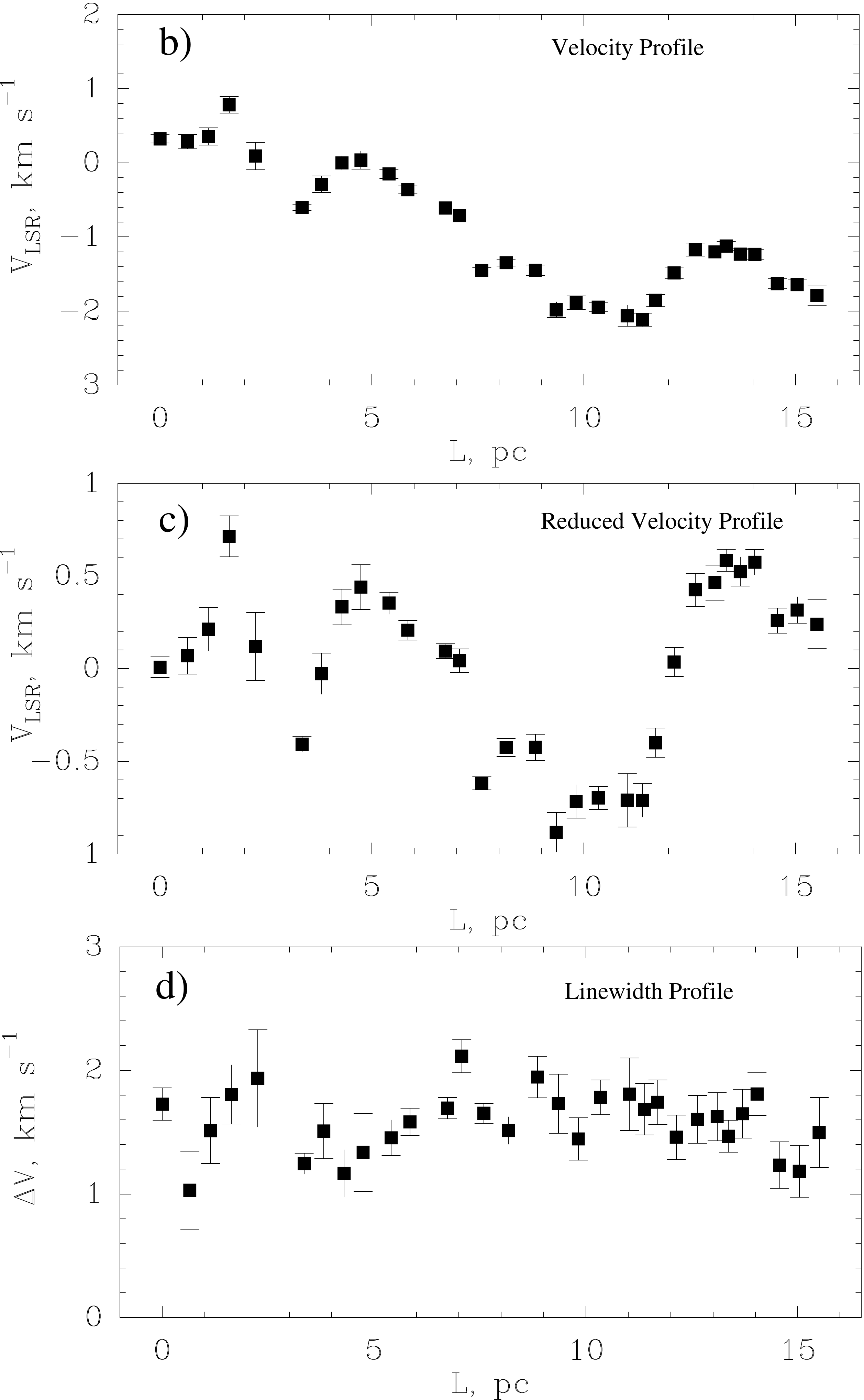}
\caption{a) FUGIN $^{13}$CO(1--0) map of intensity (moment-0) in the direction of Mon~R1. The molecular emission is integrated over a velocity range of [$-$5, 10] km s$^{-1}$. 
Blue dots are marked toward the filamentary structure, where spectra are extracted. 
b) Velocity profile along the filament obtained from distinct positions (see blue dots in panel ``a"). 
c) Velocity profile after removing the linear gradient from profile shown in panel ``b". 
d) Linewidth profile (average linewidth $\sim$1.5 km s$^{-1}$). In panels ``b", ``c", and ``d", the x-axis represents the physical length.}
\label{fg9}
\end{figure*}

Figure~\ref{fg9}a shows the integrated intensity (or moment-0) map of $^{13}$CO, where 31 distinct positions toward the filament are highlighted by filled blue circles (radius $\sim$ 10$''$ each). 
In order to examine the velocity variation of the gas toward the filament, we have extracted the averaged spectra toward each circle marked in Figure~\ref{fg9}a.
These positions are chosen where the $^{13}$CO integrated intensities are locally maximum than their surroundings. 
The observed spectra were then fitted with the Gaussian profile(s), and the spectral parameters (e.g., radial velocity (V$_\mathrm{lsr}$) and linewidth ($\Delta$V)) are derived. 
The velocity profile (i.e., velocity vs length along major axis) of the filament is shown in Figure~\ref{fg9}b, which suggests that the filament has a maximum velocity variation of $\sim$3 km s$^{-1}$. The velocity profile also shows an oscillatory pattern (with a period of $\sim$3-4 pc), which becomes more prominent when we remove the linear gradient from the profile. 
There are two prominent velocity peaks separated by $\sim$8--9 pc (see Figures~\ref{fg9}b and~\ref{fg9}c), which are associated with the central and western parts of the filament. The negative and positive velocity gradients between these peaks could be connected with different inclinations of the different parts of the filament. 
The velocity profile after removing the linear gradient is shown in Figure~\ref{fg9}c. 
There is a clear linear velocity gradient along the central part of the filament ($\sim$0.4 km s$^{-1}$ pc$^{-1}$) and also velocity gradients ($\sim$0.6 km s$^{-1}$ pc$^{-1}$) at the ends, which are probably associated with star-forming regions.
The linewidth ($\Delta$V) profile is shown in Figure~\ref{fg9}d, where one can obtain the average linewidth to be $\sim$1.5 km s$^{-1}$.
The significance of these results concerning the filamentary fragmentation is discussed in Section~\ref{sec:disc}.
\subsection{Distribution of young stellar objects}
\label{sec:yso}
\begin{figure*}
\epsscale{1}
\plotone{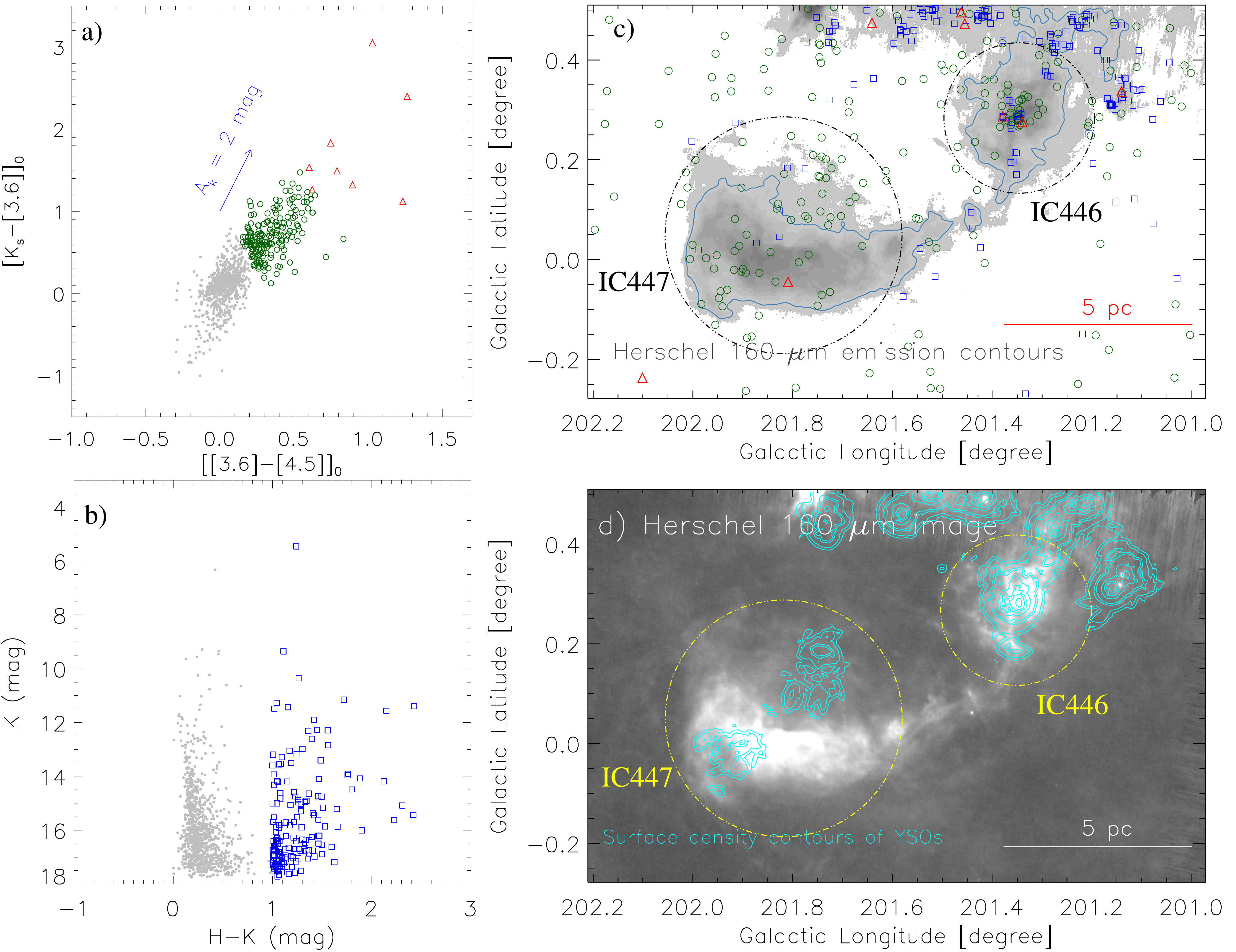}
\caption{a) Dereddened color-color ($[[3.6]-[4.5]]_{0}$ vs $[K-[3.6]]_{0}$) diagram of point-like objects toward Mon~R1. Red triangles represent Class~I YSOs, while green circles 
indicate Class~II YSOs (see text for more details). Following the extinction law given in \citet{flaherty07}, an extinction vector ($A_{K}$= 2 mag) is drawn in the figure. 
b) Color-magnitude ($H-K$ vs $K$) diagram of point-like sources. The color excess sources are highlighted by blue squares. 
c) Overlay of the selected YSOs (from panels ``a" and ``b") on the {\it Herschel} 160 $\mu$m gray-scale filled contour map. The contour values are 0.8\%, 1.5\%, 2\%, 2.5\%, 3\%, 4\%, 5\%, 7\%, 9\%, 15\%, 30\%, 50\%, 70\%, 90\%, 95\% of the peak value (i.e., 1.68 Jy pixel$^{-1}$). A contour (in sky-blue color) with a level of 0.02 Jy pixel$^{-1}$ is also displayed in the figure. d) Overlay of the surface density contours (in cyan) of YSOs on the {\it Herschel} continuum map at 160 $\mu$m. The surface density contours are presented with the levels of 4, 5, 8, 10, 20, 35, 55, and 125 YSOs pc$^{-2}$.
In panels ``c" and ``d", a scale bar corresponding to 5 pc is shown, and big dotted circles represent the extent of IC~446 and IC~447. 
In panels ``a" and ``b", sources with photospheric emission are shown by dots (in grey). Due to a large number of these sources, a small fraction is randomly displayed in panels ``a" and ``b".} 
\label{fg7}
\end{figure*}

In a given star-forming region, star formation activity is often traced by the distribution of YSOs.
The population of YSOs toward the R-association is identified using the color-color and color-magnitude diagrams (see Figures~\ref{fg7}a and~\ref{fg7}b). 
Figure~\ref{fg7}a shows the dereddened color-color diagram (i.e., $[[3.6]-[4.5]]_{0}$ vs $[K-[3.6]]_{0}$) of point-like objects extracted from 
the Warm-{\it Spitzer} Glimpse360\footnote[4]{http://www.astro.wisc.edu/sirtf/glimpse360/} \citep{whitney11} survey. 
In the dereddened color-color diagram, triangles and circles represent Class~I and Class~II YSOs, respectively. Following the conditions given in \citet{gutermuth09}, we identified a total of 201 YSOs (i.e., 8 Class~I and 193 Class~II YSOs) in our selected target area. 

Additional sources with color-excess emission are also obtained from the examination of the NIR color-magnitude ($H-K$ vs $K$) diagram (see Figure~\ref{fg7}b). 
Here we obtained the photometric data from the 2MASS and UKIDSS-GPS surveys. 
In order to obtain the credible photometric magnitudes of point-like sources from the 2MASS and UKIDSS-GPS surveys, we selected those sources which have photometric uncertainties of $\sigma$ $<$ 0.1 mag in {\it H} and {\it K} bands \citep[see also][for more details]{dewangan17a}. Following a color condition of $H-K >$ 1 mag, we obtain a total of 182 YSO candidates in our selected area (see blue squares in Figure~\ref{fg7}b). 
This color condition is decided based on the color-magnitude analysis of a nearby control field. 

Altogether, a total sum of 370 YSO candidates is selected after counting the common YSOs traced in Figures~\ref{fg7}a and~\ref{fg7}b. 
In Figure~\ref{fg7}c, the positions of these selected YSOs are overlaid on the {\it Herschel} 160 $\mu$m continuum image of Mon~R1. 
We have overplotted the {\it Herschel} 160 $\mu$m continuum emission contour to indicate the location of the filament. In order to study the concentration of YSOs in our selected area, we have carried out the nearest neighbor (NN) surface density analysis of all the selected YSOs \citep[see e.g.,][]{casertano85,gutermuth09,bressert10,dewangan17a}. Adopting similar procedures as highlighted in \citet{dewangan17a}, 
the surface density map of YSOs has been generated using a 15$\arcsec$ grid and 6 NN at a distance of 760 pc. 

In Figure~\ref{fg7}d, we overlay the surface density contours (in cyan) of YSOs on the {\it Herschel} 160 $\mu$m image. 
The surface density contours of YSOs are shown with the levels of 4, 5, 8, 10, 20, 35, 55, and 125 YSOs pc$^{-2}$, tracing the zones of star formation in our selected site. 
The YSOs are clustered mainly toward the ends of the filament (i.e., IC~446 and IC~447). 
In the direction of IC~446, we find higher surface density of YSOs (i.e., 4--125 YSOs pc$^{-2}$), while the surface density values are found to be 4-12 YSOs pc$^{-2}$ toward IC~447.
This particular analysis suggests that star formation activity is more intense toward IC~446 compared to IC~447 in the filament. 
It is also noted that the site IC~447 hosts several evolved massive B-type stars.

Most recent narrow-band H$\alpha$ and [S\,{\sc ii}] imaging observations of Mon R1 region (including sites NGC 2245, NGC 2247, and IC~446) by \citet[][submitted to MNRAS]{Movsessian20} show 
the presence of at least 4 Herbig-Haro (HH) objects toward the filament (see Figures~2 and~4 in their paper). Three out of them (i.e., HH 1202A, HH 1202B, and HH 1202C) are observed 
toward the site IC~446, while the remaining fourth one (i.e., HH 1203) is seen near the junction of the central and western parts of the filament. 
The detection of HH objects also favours high activity of star formation in the filament.
\section{Discussion}
\label{sec:disc}
The presence of a filament itself represents first step toward the formation of embedded cores and stars. However, second and major step is gravitational fragmentation which leads to the fabrication of self-gravitating cores \citep{andre10,andre14}. In this context, our observational results can be compared with the existing theoretical models of filamentary fragmentation and collapse. 
\subsection{Mon~R1 filament: a supercritical filament}
\label{sec:supercritical}
In the literature, gravitational instability of a filament is defined by its critical line mass parameter.
One can calculate the critical line mass of a filament (M$_\mathrm{l,cri}$) by considering it as an infinite and non-magnetized isothermal cylinder, which is in equilibrium between gravitational and thermal pressures. M$_\mathrm{l,cri}$ is a function of the kinetic temperature ($T_{k}$) of the gas only, and can be expressed as follows \citep{ostriker64}:

\begin{equation}
M_\mathrm{l,cri} = \frac{2c_{s}^{2}}{G} \simeq \frac{16M_{\odot}}{pc}\left(\frac{T_{k}}{10K}\right)
\label{eq2}
\end{equation}

where $c_{\rm s}=\sqrt\frac{kT_{k}}{\mu m_{H}}$ is the isothermal speed of sound at $T_{k}$, and {\it G} is the universal gravitational constant. 
Considering $T_{k}$ = 10~K and the mean molecular weight $\mu$ = 2.33, the value of $c_{\rm s}$ comes out to be $\sim$0.19 km s$^{-1}$.  
Thus, M$_\mathrm{l,cri}$ is estimated to be $\sim$24--32 M$_{\odot}$ pc$^{-1}$ for a temperature range of 15--20 K. 
In this work, the value of M$_\mathrm{l,obs}$ is determined to be $\sim$105 M$_{\odot}$ pc$^{-1}$ (see Section~\ref{sec:tmcd}), which greatly exceeds the value of M$_\mathrm{l,cri}$.

\citet{inutsuka92} found that the isothermal filaments with M$_\mathrm{l,obs} > $ M$_\mathrm{l,cri}$ become unstable to axisymmetric perturbation and collapse radially toward their axis. Such filaments are classified as thermally {\it supercritical}, which often found to nurture the prestellar cores and embedded YSOs \citep[e.g.,][]{andre10,dewangan17c}. However, the isothermal filaments which follow the condition M$_\mathrm{l,obs} < $ M$_\mathrm{l,cri}$ are designated as {\it subcritical} filaments, and show the deficiency of star formation activity \citep[e.g.,][]{andre10,dewangan17c}.
Following the above statements, Mon~R1 filament is found to be thermally {\it supercritical} in nature. The presence of the {\it Herschel} clumps/sub-filaments and groups of YSOs toward the thermally {\it supercritical} filament in Mon~R1 suggests that it is more prone to fragment and able to collapse into cores due to gravitational instability \citep{andre10}.

There is a possibility that the non-thermal microturbulent gas motions can provide extra support to the filament against gravitational collapse as these motions are associated with the gas turbulence \citep[e.g.,][]{hacar11}. 
This additional support increases the M$_\mathrm{l,cri}$ of the filament. The new line mass is termed as virial line mass (M$_\mathrm{l,vir}$) and can be estimated by replacing $c_{s}$ with the effective sound speed i.e., $c_{\rm s, eff} = \sqrt {c^{2}_{\rm s} + \sigma^{2}_{\rm NT}}$ in Eq.~\ref{eq2} \citep[e.g.,][]{kainulainen16,dewangan19}.
In the expression of $c_{\rm s, eff}$, $\sigma_{\rm NT}$ represents the non-thermal velocity dispersion that can be expressed as \citep[e.g.,][]{dewangan18}

\begin{equation}
\sigma_{\rm NT} = \sqrt{\frac{\Delta V^2}{8\ln 2}-\frac{k T_{k}}{29 m_H}} = \sqrt{\frac{\Delta V^2}{8\ln 2}-\sigma_{\rm T}^{2}} 
\label{sigmanonthermal}
\end{equation}

where $\Delta V$ is the observed linewidth of $^{13}$CO spectra, and $\sigma_{\rm T}=\sqrt{\frac{k T_{k}}{29 m_H}}$ is the thermal broadening for $^{13}$CO at 
the gas kinetic temperature (T$_{k}$). 

In this work, the linewidth of $^{13}$CO spectra is observed to be $\sim$1.5 km s$^{-1}$ (see Figure~\ref{fg9}d), which leads the value of M$_\mathrm{l,vir}$ $\sim$211--219 M$_{\odot}$ pc$^{-1}$ for a temperature range of $\sim$15--20 K. The values of M$_\mathrm{l,vir}$ and M$_\mathrm{l,obs}$ are close to each other within a factor of $\sim$2.
Mach number, defined as the ratio of $\sigma_{\rm NT}$/$c_{s}$, is determined in a range of 2.4--2.7 for the filament, which suggests the supersonic nature of the filament. 

\subsection{Age Estimates}
\label{sec:age}
Several high column density regions (i.e., fragments) are evident in the {\it Herschel} column density map, indicating the fragmentation of Mon R1 filament \citep[e.g.,][]{kainulainen16}. Due to the curved morphology, the elongated filament is divided into three parts (see Section~\ref{sec:tmcd}), which can also be traced in the molecular maps of CO (see Figures~\ref{fg2}c and~\ref{fg2}d). The eastern part containing the site IC~447 shows a separate filamentary branch in CO, 
while the central and western parts show linear geometries. Considering length of the central sub-filament (i.e., $\sim$5 pc) and the western sub-filament (i.e., $\sim$4 pc), the value of $M_\mathrm{l,obs}$ is computed to be $\sim$64 and $\sim$91 M$_{\odot}$ pc$^{-1}$, respectively (see Section~\ref{sec:tmcd}).

One can estimate a timescale at which a filament grows via accretion and reaches its M$_\mathrm{l,cri}$ value. This timescale is referred to as critical timescale ($\tau_{\rm cri}$), which represents the lower age limit of the filament \citep{clarke16}. They showed that $\tau_{\rm cri}$ is a function of fragmentation length scale (i.e., fragment/core separation, $\lambda_{\rm core}$), which can be expressed as follows \citep{Williams18}: 

\begin{equation}
\tau_{\rm age} \geq \tau_{\rm cri} \simeq \frac{\lambda_{\rm core}}{2c_{\rm s}}
\label{tau}
\end{equation}

We have estimated $\lambda_{\rm core}$ along the central and western parts of Mon R1 filament, which is $\sim$0.5 pc and $\sim$0.2 pc, respectively (see Figures~\ref{fg4}b and~\ref{fg9}c). For $c_{\rm s}$ $\sim$0.23 km s$^{-1}$ (at $T_{k}$ = 15 K), these core separation values lead $\tau_{\rm cri}$ to be $\sim$1.1 Myr and $\sim$0.4 Myr for the central and western parts, respectively. 
The corresponding average accretion rate during the assembly of filament can also be calculated using $\dot{M}=\frac{M_{\rm l,cri}}{\tau_{\rm cri}}$ 
\citep[e.g.,][]{clarke16,Williams18}. 
Thus, considering $M_{\rm l,cri}$ $\sim$24 M$_{\odot}$ pc$^{-1}$ (at $T_{k}$ = 15 K) for both the central and western parts of the filament, 
the average accretion rate experienced during the growth of these parts of filament is $\sim$22 and $\sim$60 M$_{\odot}$ pc$^{-1}$ Myr$^{-1}$, respectively. 

If we assume that the accretion rate remained constant over the filament's lifetime, then the age of the filament can be estimated 
using $\tau_{\rm age}=\frac{M_{\rm l,obs}}{\dot{M}}$ \citep[e.g.,][]{Williams18}. 
This exercise yields the values of $\tau_{\rm age}$ to be $\sim$3 and $\sim$1.5 Myr for the central and western parts of the filament, respectively. 
Therefore, the elapsed time since the central and western parts of the filament become critical, is $\sim$1.9 and $\sim$1.1 Myr, respectively.
However, considering the turbulence into account $c_{\rm s}$ can be replaced by $c_{\rm s,eff}$ in Eq.~\ref{tau}. Adopting $T_{k}$ = 15 K and $\Delta$V$\sim$ 1.5 km s$^{-1}$ (i.e., FWHM($^{13}$CO)), the corresponding critical age for the central and western parts becomes $\sim$0.4 and $\sim$0.2 Myr, respectively. 
The projection effects are not taken into account in the estimations of $\lambda_{\rm core}$. Therefore, all these calculations give lower limits on the derived values of ages.

We use the $^{13}$CO line data to trace the velocity profile of the filament. Figures~\ref{fg9}b and~\ref{fg9}c show an ordered oscillatory pattern in velocity (with a period of $\sim$3--4 pc) along the major axis of the filament. The previous study of L1517 dark cloud by \citet{hacar11} showed a similar trend in the velocity profile. They also performed the fragmentation model for L1517 dark cloud and found that the velocity centroids peak at the locations of fragments, indicating gas motion associated with the fragments. Similar trends have also been observed in the other studies \citep[e.g.,][]{hacar16,kainulainen16,Lu18,dewangan19}.
These observational signatures of the fragmentation in filaments are also consistent in Mon~R1 filament. However, to study the nature of fragments in Mon~R1 filament, one requires detail modelling of filament using the information of so far unknown properties (i.e., orientation, geometry and strength of the magnetic field). 
Also, the availability of dense gas tracers in the filaments (e.g., HNC (1--0); \citet{Jackson10}, HCO$^{+}$ (3--2); \citet{Zernickel13}, N$_{2}$H$^{+}$ (1--0); \citet{Beuther15}, NH$_{3}$ (1--1, 2--2); \citet{Lu18}, CS (2--1); \citet{dewangan19}) can further constrain the velocity field which is the observational limitation of our current work.
\subsection{Star Formation Scenario}
Two different collapse modes of a filament have been discussed by \citet{pon12} in their study. 
They suggest that the interior of the filament collapses homologously, while its ends are preferentially given more momentum thus tend to collapse faster. 
In the case of the homologous collapse, density remains uniform and all the regions have a similar collapse timescale. 
The analytical study of an elongated but finite-sized filament suggests that the collapse timescale depends upon the relative position along the filament 
and it is shorter at the edges of the filament \citep{bastien83,pon11}. 
For a filament having an initial aspect ratio ($A$) = 5, the collapse timescale at its edges is observed 2--3 times less compared to the central regions, 
while those with higher aspect ratio collapse much faster at the ends \citep[see Figures~5 and~6 in][]{pon11}. 
Here $A$ is defined as $\frac{L}{R}$, where 2$L$ is the length of the filament and $R$ is the radius of it \citep[e.g.,][]{bastien83,toala12,pon12}. 
The faster collapse at the edges of the filament is supported by the high preferential acceleration of the gas, which leads to density enhancement there (see Section~2.2 in \citet{pon12}). 
Interestingly, the western part of the filament is found to have higher accretion rate (i.e., $\sim$60 M$_{\odot}$ pc$^{-1}$ Myr$^{-1}$) compared to the central part (i.e., $\sim$22 M$_{\odot}$ pc$^{-1}$ Myr$^{-1}$; see Section~\ref{sec:age}). 
The homologous collapse is observed if $A$ is less than 5, while the end-dominated collapse is dominant if $A\ge$ 5 \citep{pon11,pon12}. 
Using the $^{13}$CO moment-0 map (see Figure~\ref{fg2}d), the observed aspect ratio of the Mon~R1 filament is estimated to be $\sim$11.5 (i.e., for length $\sim$14 pc and diameter $\sim$1.2 pc).
This value is larger than 5, suggesting the faster collapse at its ends.

Observationally, we find embedded {\it Herschel} high column density regions/fragments and the periodic oscillation in the velocity profile of the filament, tracing the ongoing fragmentation 
and formation of the precursor of potential gravitationally bound cores. 
Furthermore, clusters of YSOs, sites IC~446 and IC~447, and massive B-type stars are observed at the ends of the filament.
It suggests the intense star formation activities at the filament ends compared to its central part, implying that the local collapse had occurred faster at the filament ends.  
Altogether, our observational results are consistent with the ``end dominated collapse"' model of star formation in Mon~R1 filament. The observational examples of end dominated fragments/clumps or end dominated collapse scenario are a few in the literature
(such as, NGC 6334 \citep{Zernickel13}; IRDC~18223 \citep{Beuther15}; Musca cloud \citep{hacar16,kainulainen16}; S242 \citep{dewangan17c,dewangan19}; G341.244$-$00.265 \citep{Yu19}). 
Out of these, the filament in S242 shows the evolved phase of filamentary fragmentation as its one end hosts the S242 H{\sc ii} region, which might have formed by the process of ``end dominated collapse". Rest of the mentioned filaments are caught in their very early phases of fragmentation as they show dense clumps along their major axis. 
At least one of the ends of a few filaments show higher velocity gradients compared to central region (e.g., 0.5 km s$^{-1}$ pc$^{-1}$ toward the southern end of Musca cloud \citep{hacar16,kainulainen16}; 1 km s$^{-1}$ pc$^{-1}$ toward both the ends of S242 \citep{dewangan17c,dewangan19}; 0.6--0.8 km s$^{-1}$ pc$^{-1}$ toward one end of G341.244$-$00.265 \citep{Yu19}). In contradiction, the filament in NGC~6334 displays the velocity gradient from the ends toward the centre \citep{Zernickel13}, and IRDC~18223 filament shows a transverse velocity gradient at one of its ends instead of showing velocity gradient along the major axis.

In comparison to the listed filaments, Mon R1 filament can be treated as the best example of a more evolved stage of the filamentary fragmentation, which reveals end dominated collapse thus forming the known massive star-forming sites at the ends. Our observational analysis shows the clear linear velocity gradient along the central part of the filament ($\sim$0.4 km s$^{-1}$ pc$^{-1}$), and relatively higher velocity gradients ($\sim$0.6 km s$^{-1}$ pc$^{-1}$) at the ends, which are probably associated with star-forming regions. 
\section{Summary and Conclusions}
\label{sec:conc}
This paper aims to understand the physical processes governing the star formation activities in Mon~R1 filament. In this context, a multi-wavelength observational approach has 
been employed in Mon~R1 complex. Our major observational results are pointed below:\\
$\bullet$ The optical and infrared images show the existence of a filament (length $\sim$14 pc, mass $\sim$1465 M$_{\odot}$) in Mon~R1, which is the most prominent feature in the {\it Herschel} column density map.\\
$\bullet$ Using the $^{12}$CO and $^{13}$CO line data, two molecular clouds at [$-$5, 1] and [2, 10] km s$^{-1}$ are traced toward our selected target area. 
The molecular gas associated with the filament is studied in a velocity range of [$-$5, 1] km s$^{-1}$, which is found to be isolated from another diffuse cloud at [2, 10] km s$^{-1}$. 
The analysis of molecular gas has enabled us to depict the exact boundaries of the filamentary cloud.\\
$\bullet$ The filament is found to contain two previously known sites IC~446 and IC~447 at its opposite ends. Several massive B-type stars are identified toward the site IC~447, 
while a massive YSO (i.e., VY~Mon) is observed toward the site IC~446.\\
$\bullet$ The extended temperature features (at $T_\mathrm{d}\sim$ 15--21 K), PDRs, and outflow activity are observed toward both the ends of the filament, suggesting 
the ongoing star formation activity.\\ 
$\bullet$ The study of the distribution of selected YSOs reveals clustering of YSOs mainly toward the ends of the filament (i.e., IC~446 and IC~447). 
Higher surface density of YSOs (i.e., 4--125 YSOs pc$^{-2}$) is traced toward IC~446, while in the direction of IC~447, the surface density is found to be 4-12 YSOs pc$^{-2}$. 
It indicates the intense star formation activities toward IC~446 compared to IC~447.\\ 
$\bullet$ 
Using the {\it Herschel} column density map, the total estimated mass of the filament (length $\sim$14 pc) is $\sim$1465 M$_{\odot}$, which yields the observed line mass 
(M$_{l,obs}$) to be $\sim$105 M$_{\odot}$ pc$^{-1}$. The critical line mass of the filament (M$_{l,cri}$) is computed to be $\sim$24--32 M$_{\odot}$ pc$^{-1}$ for a temperature range of $\sim$15--20 K. It implies that M$_{l,obs}$ is $\sim$3--4 times more than M$_{l,cri}$, suggesting the existence of a thermally {\it supercritical} filament in Mon~R1.\\  
$\bullet$ The western part of the filament takes about 0.4 Myr to grow its line mass by its critical value with an 
average accretion rate of $\sim$60 M$_{\odot}$ pc$^{-1}$ Myr$^{-1}$. In the case of the central part of the filament, this critical timescale and the accretion rate are estimated to be $\sim$1.1 Myr 
and $\sim$22 M$_{\odot}$ pc$^{-1}$ Myr$^{-1}$, respectively.\\
$\bullet$ The velocity and linewidth profiles of the filament show an oscillatory pattern with a periodicity of 3--4 pc along the direction of its major axis. 
The velocity gradient along the central part of the filament is found to be ($\sim$0.4 km s$^{-1}$ pc$^{-1}$), however the relatively higher velocity gradients ($\sim$0.6 km s$^{-1}$ pc$^{-1}$) are observed at both the ends of the filament.\\
$\bullet$ The aspect ratio ({\it A}) of the filament is found to be greater than 5, suggesting the end dominated collapse mode is dominated over the homologous collapse mode.\\

Taken together, all the observational outcomes, the fragments distributed toward both the ends of the filament are found to undergo a faster collapse compared to its central part, showing the presence of massive star-forming regions at its opposite ends. Overall, the elongated filament in Mon~R1 is a promising candidate favouring the ``end-dominated collapse" model of 
star formation as discussed by Pon et al. (2011, 2012).
\acknowledgments  
We thank the anonymous reviewer for several useful comments and suggestions. 
The research work at Physical Research Laboratory is funded by the 
Department of Space, Government of India. 
This work is based on data obtained as part of the UKIRT Infrared Deep Sky Survey. 
This publication makes use of data products from the Two Micron All Sky Survey, which is a joint project of the University of Massachusetts and the Infrared 
Processing and Analysis Center/California Institute of Technology, funded by NASA and NSF.
This work is based [in part] on observations made with the {\it Spitzer} Space Telescope, which is operated by the Jet Propulsion Laboratory, California Institute of Technology under a contract with NASA. This publication makes use of data from FUGIN, FOREST Unbiased Galactic plane Imaging survey with the Nobeyama 45-m telescope, a legacy project 
in the Nobeyama 45-m radio telescope. 
LEP acknowledges support of the Russian Science Foundation (project 17-12-01256). 
DKO acknowledges the support of the Department of Atomic Energy, Government of India, under project No. 12-R\&D-TFR-5.02-0200.

\end{document}